\documentclass[aps,prl,twocolumn,superscriptaddress,longbibliography]{revtex4-2}
\usepackage{graphicx}
\usepackage{amsmath,amssymb}
\usepackage{bm}
\usepackage{hyperref}   
\usepackage{microtype}  
\usepackage[dvipsnames]{xcolor}

\newcommand{\jkrem}[1]{{\color{blue}}}

\begin{document}

\title{Nonmagnetic Ground State of Rutile RuO$_2$ from Diffusion Quantum Monte Carlo}

\author{Jeonghwan Ahn}
\email{kindazet@gmail.com}
\affiliation{Materials Science and Technology Division, Oak Ridge National Laboratory, Oak Ridge, Tennessee 37831, USA}
\affiliation{The Anthony J Leggett Institute for Condensed Matter Theory, Department of Physics, University of Illinois at Urbana-Champaign, Urbana, Illinois 61801, USA}
\author{Seoung-Hun Kang}
\email{shkang@kisti.re.kr}
\affiliation{Materials Science and Technology Division, Oak Ridge National Laboratory, Oak Ridge, Tennessee 37831, USA}
\affiliation{Department of Information Display, Kyung Hee University, Seoul 02447, Korea}
\affiliation{Research Center for Technology Commercialization, Korea Institute of Science and Technology Information (KISTI), Seoul 02456, Korea}
\author{Panchapakesan Ganesh} 
\email{ganeshp@ornl.gov}
\affiliation{Center for Nanophase Materials Sciences, Oak Ridge National Laboratory, Oak Ridge, Tennessee 37831, USA}
\author{Jaron T. Krogel}
\email{krogeljt@ornl.gov}
\affiliation{Materials Science and Technology Division, Oak Ridge National Laboratory, Oak Ridge, Tennessee 37831, USA}

\date{\today}


\begin{abstract}
Rutile RuO$_2$ has been proposed as an altermagnet, but its bulk magnetic ground state is still under debate because density-functional calculations give conflicting predictions. Using fixed-node diffusion quantum Monte Carlo, we find that stoichiometric bulk RuO$_2$ is nonmagnetic in the pristine structure, lying 23(9) meV per formula unit below the lowest antiferromagnetic state considered. A 3$\%$ compressive strain instead stabilizes antiferromagnetism, placing RuO$_2$ near a strain-tunable magnetic instability and helping reconcile apparently conflicting experimental reports.

\end{abstract}

\maketitle

\begin{figure}[t]
\centering
\includegraphics[width=\columnwidth]{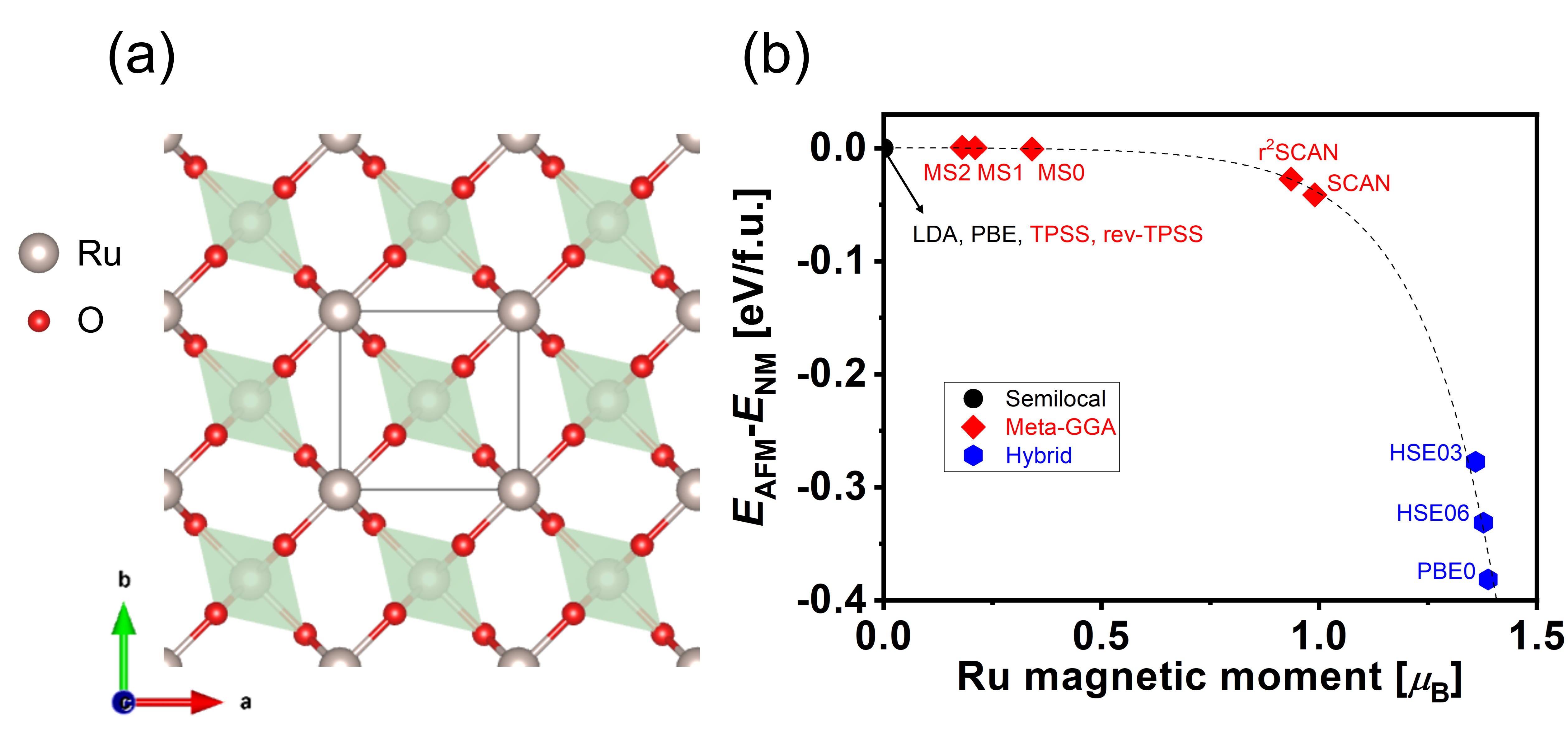}
\caption{(a) Crystal structure of rutile RuO$_2$ (RuO$_6$ octahedra). (b) DFT magnetic ordering energy $E_{\mathrm{AFM}}-E_{\mathrm{NM}}$ versus Ru local moment across representative functionals; dotted line is a guide to the eye.}
\label{fig:inv_Jacobs_ladder}
\end{figure}

Altermagnetism has recently emerged as a symmetry-protected paradigm in magnetism that bridges key properties of ferromagnetic (FM) and antiferromagnetic (AFM) phases through compensated magnetic structures that nevertheless host spin-split electronic responses~\cite{vsmejkal2022emerging,vsmejkal2022beyond}. Altermagnets thus offer routes to spintronic functionalities that can circumvent fundamental constraints inherent to conventional FM and AFM platforms~\cite{vsmejkal2023chiral,amin2024nanoscale,bai2024altermagnetism,song2025altermagnets,jeong2025magnetic}. In rutile RuO$_2$, where Ru is octahedrally coordinated by O to form edge-sharing RuO$_6$ chains linked by corner sharing [Fig.~\ref{fig:inv_Jacobs_ladder}(a)], the crystal symmetry satisfies these requirements, and RuO$_2$ has consequently been regarded as a compelling candidate to realize altermagnetic order~\cite{vsmejkal2022emerging,vsmejkal2022beyond,feng2022anomalous,bose2022tilted,fedchenko2024observation}. Establishing whether stoichiometric RuO$_2$ hosts intrinsic magnetism is therefore essential for assessing the feasibility of altermagnetism in real materials.

Despite this promise, the magnetic ground state of RuO$_2$ remains unsettled. Early x-ray scattering~\cite{zhu2019anomalous} and neutron diffraction~\cite{berlijn2017itinerant} reported AFM order, and subsequent experiments~\cite{feng2022anomalous,bose2022tilted,fedchenko2024observation} together with DFT calculations with Hubbard-$U$ corrections~\cite{vsmejkal2020crystal,vsmejkal2022beyond} supported spin-polarized states. In contrast, recent muon spin rotation ($\mu$SR) measurements~\cite{hiraishi2024nonmagnetic,kessler2024absence} failed to detect static magnetic moments in bulk or epitaxial samples, while complementary bulk-sensitive probes also find an electronic response consistent with a paramagnetic metal and place stringent constraints on any intrinsic spin polarization in stoichiometric RuO$_2$~\cite{wenzel2025fermi,wu2025fermi}. This discrepancy has sharpened the debate on whether altermagnetic-like features are intrinsic to stoichiometric RuO$_2$ or instead driven by extrinsic tuning such as strain, disorder, or doping~\cite{smolyanyuk2024fragility,wickramaratne2025effects,forte2025strain}.

A direct and quantitatively reliable determination of magnetic energetics is thus required. On the theory side, DFT+$U$ has been the primary tool for examining magnetic states, yet it is not a controlled framework for itinerant $4d$ metals: predicted moments can depend strongly on both the exchange-correlation functional and the value of $U$, and there is no clear \emph{a priori} furst-principles criterion for selecting a physically reasonable $U$ in this regime~\cite{anisimov1991band,petukhov2003correlated,himmetoglu2014hubbard,petukhov2003correlated,ylvisaker2009anisotropy,ryee2018effect,smolyanyuk2024fragility,yumnam2025constraints}. Meta-GGA functionals such as SCAN can likewise spuriously stabilize magnetism and overestimate magnetic energy scales in itinerant metals~\cite{ekholm2018assessing,fu2019density,mejia2019analysis,Tran2020shortcomings,desmarais2025meta,fu2018applicability}. Taken together, these considerations call for a many-body benchmark that can determine whether stoichiometric RuO$_{2}$ is intrinsically nonmagnetic or magnetically ordered.

Figure~\ref{fig:inv_Jacobs_ladder}(b) highlights a key limitation of mean-field energetics for RuO$_2$.
The energy difference between the NM and AFM solutions, which we define as the magnetic ordering energy $E_\text{AFM}-E_\text{NM}$, varies substantially between exchange-correlation functionals and closely tracks the Ru local moment stabilized at the mean-field level. 
Here and throughout, the Ru local moment is estimated from the spin-density projection within the Ru-centered local region for both DFT and DMC calculations, as described in the Supplemental Material~\cite{SI}.
Climbing Jacob’s ladder from semilocal to meta-GGA and hybrid functionals progressively stabilizes AFM order and increases the associated Ru moments.
Within the tunable DFT+$U$ and PBE0($\omega$) families, both the magnetic ordering energy and the Ru local moment likewise increase as the corresponding tuning parameters are raised (See Fig. 1 of the Supplemental Material~\cite{SI}). 
A similar trend has also been reported in elemental itinerant magnets~\cite{ekholm2018assessing,fu2018applicability,fu2019density,mejia2019analysis,Tran2020shortcomings,desmarais2025meta}, indicating that this behavior is not unique to RuO$_2$ but reflects a broader difficulty of approximate density functionals in itinerant magnetic regimes. 
In RuO$_2$, however, this sensitivity becomes especially consequential because the relevant magnetization and energy scales are both small. 
As a result, modest changes in functional choice or tuning parameters can shift the Ru moment and the NM-AFM energy balance by amounts comparable to the entire disputed energy scale.

More importantly, the progression to higher-rung functionals does not converge toward the experimentally constrained regime. Recent $\mu$SR measurements support a nonmagnetic metallic ground state~\cite{hiraishi2024nonmagnetic,kessler2024absence}, and neutron scattering reports at most a small ordered moment of $M \approx 0.05~\mu_{\text{B}}$~\cite{berlijn2017itinerant}, whereas higher rungs yield increasingly large AFM moments and magnetic ordering energies.
Furthermore, hybrid functionals and DFT+$U$ with higher values of $U$ can predict insulating AFM states for RuO$_2$ (see Fig. 2 of the Supplemental Material~\cite{SI}), inconsistent with metallic transport~\cite{berlijn2017itinerant,zhu2019anomalous,feng2022anomalous}.
Thus, the relevant energy scale is small while the methodological spread is large, and any viable theory must also be consistent with experiments showing metallic transport with either no static magnetism ($\mu$SR)~\cite{hiraishi2024nonmagnetic,kessler2024absence} or, at most, an extremely small ordered moment in neutron measurements~\cite{berlijn2017itinerant}.
We therefore use fixed-node diffusion quantum Monte Carlo (DMC) as a variational discriminator to determine the energetically preferred magnetic state in the intrinsic limit and to assess how external tuning, such as strain, shifts the NM-AFM balance.

\begin{figure}[t]
\centering
\includegraphics[width=\columnwidth]{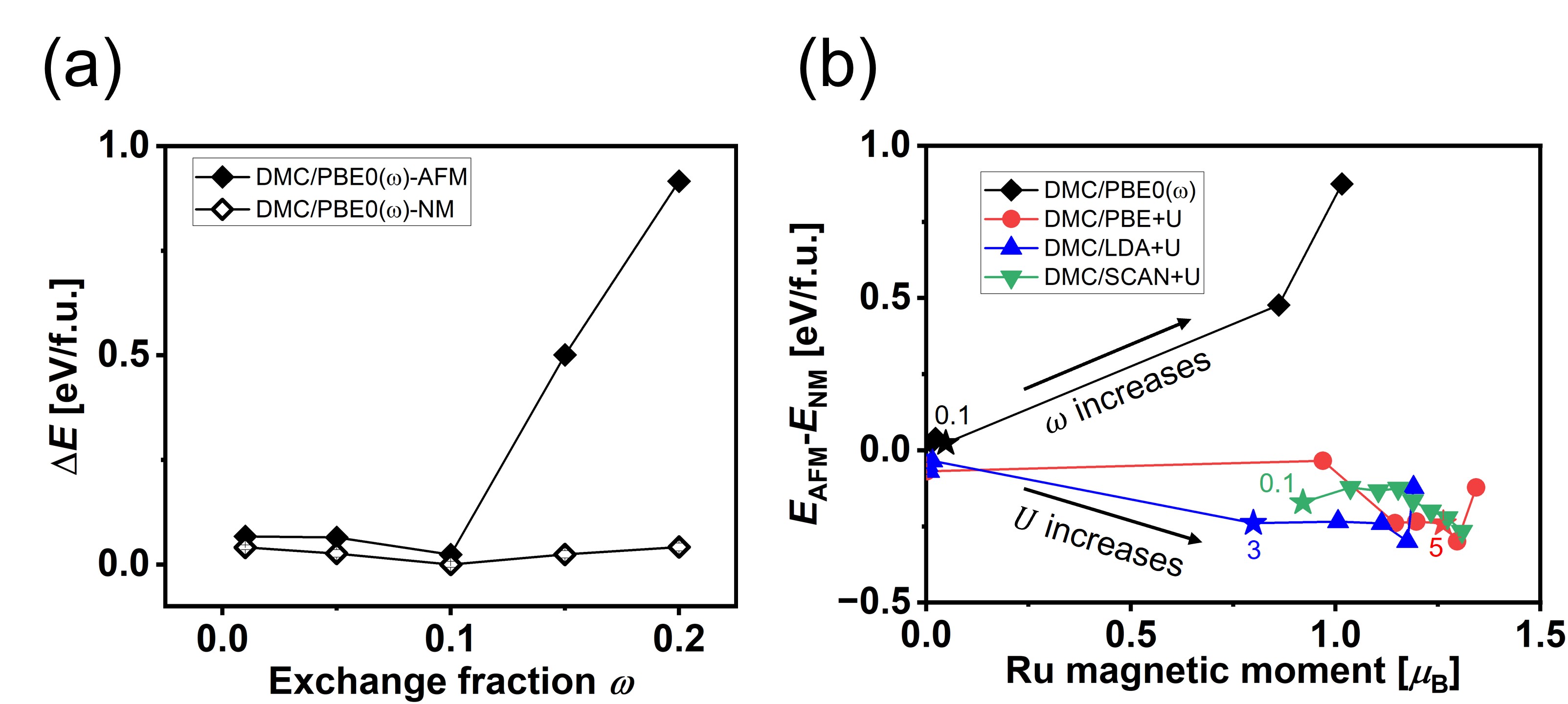}
\caption{(a) Twist-averaged fixed-node DMC energies for NM and AFM states using PBE0($\omega$) trial nodes. 
(b) DMC magnetic ordering energy $E_{\mathrm{AFM}}-E_{\mathrm{NM}}$ versus Ru local moment estimated with DMC using PBE0($\omega$) and PBE+$U$. Stars denote the variationally optimal point within each functional class, representing the best value of the exact-exchange fraction $\omega$ for PBE0 and the optimal Hubbard $U$ parameter for the PBE+$U$ functionals.
}
\label{fig:E_vs_U}
\end{figure}

Figure~\ref{fig:E_vs_U}(a) summarizes our fixed-node DMC determination of the magnetic ground state.
We performed fixed-node DMC calculations using QMCPACK~\cite{kim18,kent20}, and additional computational details are provided in Sec.~I of the Supplemental Material~\cite{SI}.
Because the fixed-node DMC energy depends variationally on the nodal surface, we treat $U$ (DFT+$U$) and the exact-exchange fraction $\omega$ (in PBE0) as practical tuning parameters for the trial nodes within each ansatz, enabling a direct many-body energy comparison across these trial-wave-function families. Note that in most DFT implementations of the PBE0 functional, the default choice is $\omega=0.25$~\cite{note} as there is no self-consistent approach to determine this value within DFT.

Within the PBE0($\omega$) family, the NM state exhibits a clear DMC energy minimum at $\omega=0.10$.
At this variational minimum, the NM solution lies 23(9) meV per formula unit below the lowest AFM state considered.
This preference for the NM states remains unchanged across the convergence checks performed here, as summarized in Fig. 3 of the Supplemental Material~\cite{SI}.
Importantly, $\omega$ is selected here by the fixed-node variational principle through many-body calculations rather than by fitting to any experimental observable.

Among the trial wave functions considered here, the lowest fixed-node DMC energy is obtained for the NM state constructed from PBE0($\omega=0.10$) orbitals. 
By contrast, DMC calculations based on LDA+$U$, PBE+$U$, and SCAN+$U$ trial orbitals favor AFM solutions within those respective families, although their family minima remain above the optimized PBE0 result (See Fig. 4 of the Supplemental Material~\cite{SI}). 
At the mean-field level, the variationally selected NM trial state remains metallic, whereas the large-$\omega$ and large-$U$ magnetic solutions that have higher variational energy become insulating, consistent with transport experiments~\cite{berlijn2017itinerant,zhu2019anomalous,feng2022anomalous}. 
This nonmagnetic ground-state assignment is also consistent with the absence of static moments reported by recent $\mu$SR studies~\cite{hiraishi2024nonmagnetic,kessler2024absence}. 
Taken together, these results indicate that the stoichiometric rutile RuO$_{2}$ is nonmagnetic in the pristine bulk structure.

Figure~\ref{fig:E_vs_U}(b) provides a complementary view by plotting the magnetic ordering energy $E_{\mathrm{AFM}}-E_{\mathrm{NM}}$ as a function of the Ru local moment, analogous to Fig.~\ref{fig:inv_Jacobs_ladder}(b).
While DFT+$U$ can yield a NM mean-field solution at sufficiently small $U$, the DMC minima within the $+U$ scans occur at larger $U$, where the mean-field solutions carry sizable Ru moments, and the AFM state is energetically lower (See Fig. 1 of the Supplemental Material~\cite{SI}).
In contrast, for PBE0($\omega$)-based trial functions, $E_{\mathrm{AFM}}-E_{\mathrm{NM}}$ remains positive over the entire tested range and increases with $\omega$.
This indicates that increasing the exact exchange raises the energetic cost of imposing AFM order once correlation is treated at the DMC level.
Consistently, the DMC minimum in the PBE0($\omega$) scan occurs at small $\omega$, where the mean-field solution is already nonmagnetic, and the lowest-energy DMC state is likewise nonmagnetic (See Fig. 1 of the Supplemental Material~\cite{SI}).

The contrasting trends in Fig.~\ref{fig:E_vs_U}(b) can be understood from the different ways in which DFT+$U$ and PBE0($\omega$) reshape the underlying one-particle manifold. Although the physics underlying DFT + $U$ and PBE0($\omega$) both tend to favor unpaired spins, the former favors more localization by construction, making it a relatively poorer choice for itinerant metals. 
In DFT+$U$, the on-site Coulomb term selectively enhances localization within the Ru-4$d$ subspace, which stabilizes spin polarization and drives the trial solutions toward large-moment AFM states. 
By contrast, the nonlocal exact-exchange term in PBE0 acts on the full hybridized Ru-O manifold. 
For moderate values of $\omega$, this reshapes the electronic structure of Ru-O without strongly localizing the 4$d$ states, allowing the itinerant NM solution to remain stable at the mean-field level. 
As a result, the PBE0($\omega$) family provides a trial space in which both itinerant NM and spin-polarized AFM solutions can be explored without imposing strong localization biases. 
Within this space, the fixed-node variational criterion selects the NM solution at $\omega$ = 0.10 as the lowest-energy state.

We therefore conclude that RuO$_{2}$ has a nonmagnetic metallic ground state in the pristine structure. This finding suggests that previously reported magnetic or altermagnetic-like signatures may originate from perturbations beyond the ideal bulk limit, with anisotropic strain emerging here as one concrete mechanism.

\begin{figure}[t]
\centering
\includegraphics[width=\columnwidth]{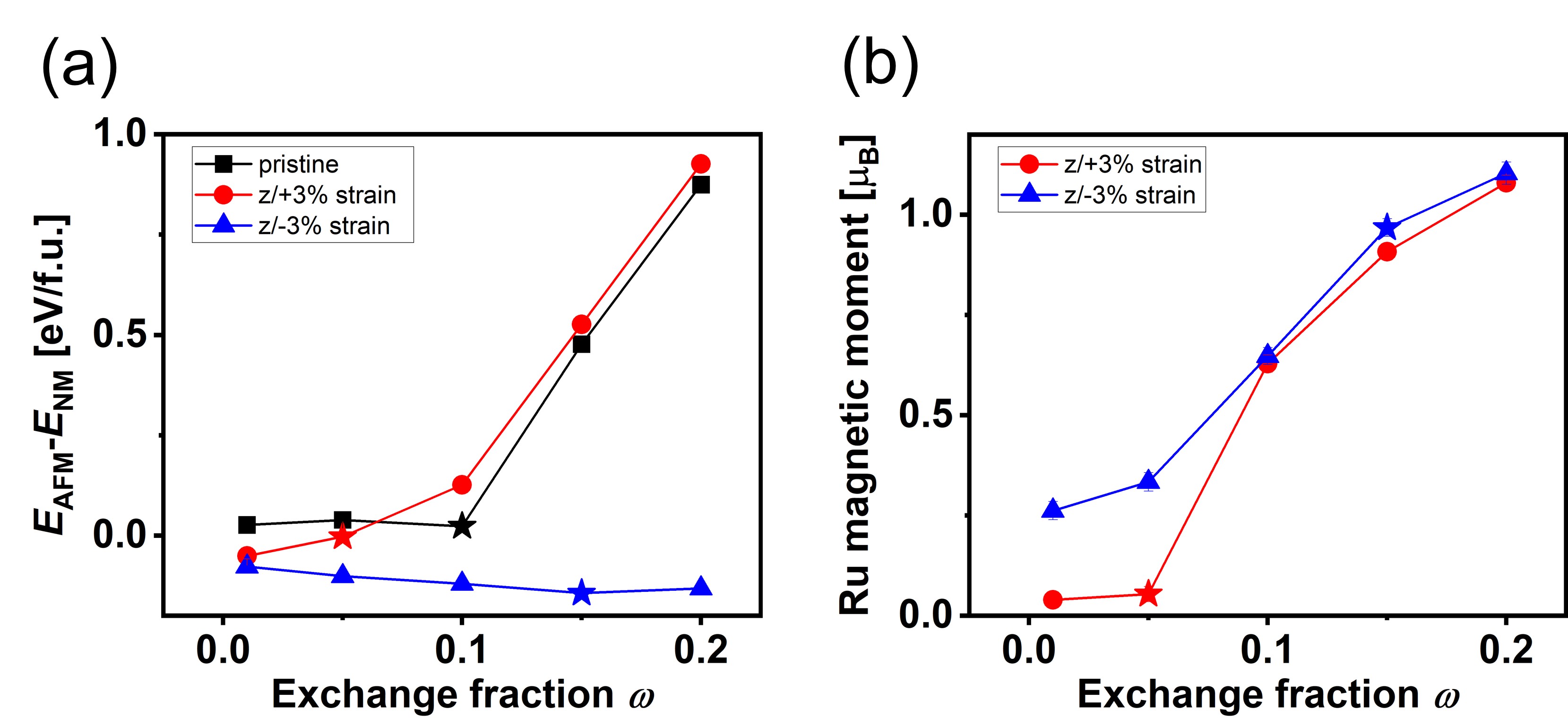}
\caption{(a) Fixed-node DMC $E_{\mathrm{AFM}}-E_{\mathrm{NM}}$ under $\pm 3\%$ $z$ strain using PBE0($\omega$) trial nodes. (b) Ru local moment versus $\omega$ for strained RuO$_2$. Stars denote the variationally optimal exact-exchange fraction $\omega$ of PBE0 for each applied strain case.}
\label{fig:strain}
\end{figure}

We next examine whether epitaxial-like strain can shift the delicate NM-AFM balance identified above. To this end, we apply $\pm 3\%$ strain along the $z$ direction while relaxing the in-plane lattice constants at the PBE level. Figure~\ref{fig:strain}(a) shows that under $3\%$ compressive strain, the magnetic ordering energy becomes negative and an AFM state is stabilized, whereas under $3\%$ tensile strain the NM state remains favored within statistical uncertainty. 
At $3\%$ compressive strain , the AFM solution carries an ordered moment of 0.97(2) $\mu_{\mathrm{B}}$ (See Fig.~\ref{fig:strain}(b)).

Consistent with the evolution of both the magnetic ordering energy and the Ru moment, the optimal $\omega$ shifts toward larger values under compression and toward smaller values under tension. 
Thus, compressive strain drives RuO$_2$ toward a more correlated magnetic regime, whereas tensile strain preserves the itinerant nonmagnetic character of the bulk state.

These results place RuO$_2$ on the verge of magnetism: the stoichiometric bulk limit remains nonmagnetic, while modest compressive distortions stabilize an AFM state with a sizable Ru moment.
This provides a natural interpretation of earlier experimental signatures associated with altermagnetic order in RuO$_{2}$~\cite{feng2022anomalous,bose2022tilted,fedchenko2024observation,zhu2019anomalous,berlijn2017itinerant}: they need not imply an intrinsically magnetic bulk ground state, but can instead arise when epitaxial strain or substrate-induced pressure pushes the system across the nearby NM-AFM boundary.

Indeed, the possibility of an altermagnetic response in RuO$_2$ continues to be actively discussed in epitaxially grown thin films~\cite{he2025evidence,jeong2025anisotropic,jeong2025metallicity,forte2025strain,brahimi2025confinement,lytvynenko2026magnetic,akashdeep2026surface}.
Strain control therefore emerges as a key knob for tuning RuO$_2$ between itinerant nonmagnetic and correlated magnetic behavior.
More broadly, this balance cannot be resolved reliably by simply adjusting $\omega$ or $U$ within DFT or DFT+$U$ alone, but instead requires the many-body variational framework provided by DMC.

\begin{figure}[t]
\centering
\includegraphics[width=\columnwidth]{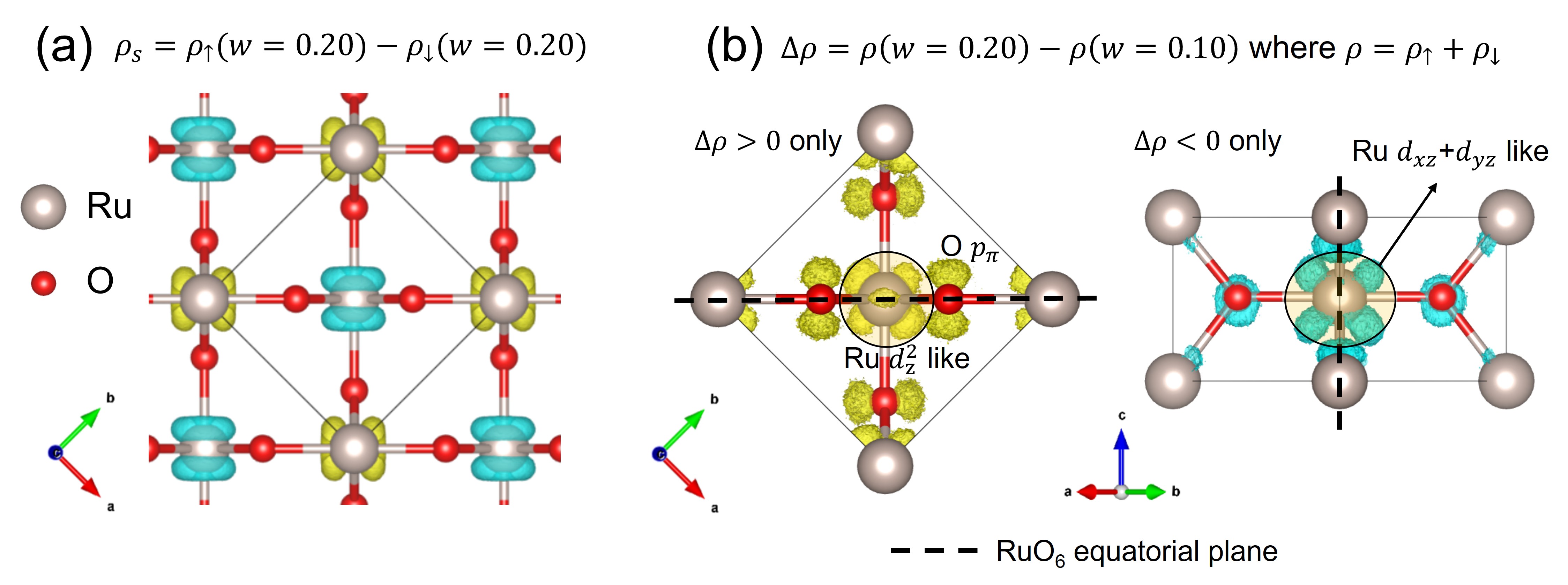}
\caption{(a) Isosurfaces of spin density distribution for AFM phase obtained from DMC based on PBE0 orbitals with $\omega = 0.20$. The isosurface levels are $\pm 2 \times 10^{-5}$ with the spin-up and spin-down electrons being represented by yellow and blue colors, respectively. (b) Isosurface of the DMC charge-density difference between calculations based on PBE0 orbitals with $\omega=0.20$ and $\omega=0.10$. Yellow indicates regions where the DMC charge density is higher for $\omega=0.20$ than for $\omega=0.10$, while blue indicates regions where it is lower. The isosurface levels are $\pm 3\times 10^{-6}$.}
\label{fig:charge_diff}
\end{figure}

Finally, we compared the variationally optimal NM state at $\omega$ = 0.10 with a representative large-moment AFM state at $\omega$ = 0.20 to identify the real-space charge and spin-density rearrangements associated with the AFM solution relative to the variationally optimal NM state.
We stress that $\omega=0.20$ is used only to provide a clear real-space contrast and does not represent the lowest AFM excitation. 
Figure~\ref{fig:charge_diff}(a) displays the DMC spin density for the AFM state based on PBE0($\omega=0.20$) trial nodes. 
The resulting pattern exhibits an altermagnetic-like texture in the Ru-O network, with sign-alternating spin lobes related by a $90^\circ$ rotation, showing that the symmetry-consistent local texture is preserved even in the presence of many-body correlations.

Figure~\ref{fig:charge_diff}(b) shows the difference between the DMC charge densities of the $\omega=0.20$ AFM and $\omega=0.10$ NM solutions. 
Yellow (blue) isosurfaces mark regions where the AFM state carries more (less) charge than the variationally optimal NM state. 
Around the Ru site, the $\omega=0.20$ AFM state exhibits enhanced density relative to the $\omega=0.10$ NM state, consistent with a $d_{z^2}$-like $e_g$ component on the left side of Fig.~\ref{fig:charge_diff}(b), accompanied by a depletion pattern with $d_{xz}$+$d_{yz}$-like $t_{2g}$ character on the right side.

On the surrounding O atoms, the AFM state shows enhanced amplitude in $p_{\pi}$-like lobes pointing perpendicular to the RuO$6$ equatorial plane, relative to the $\omega=0.10$ NM state. 
Taken together, these features indicate that increasing $\omega$ drives the trial state away from more itinerant in-plane Ru--O hybridization and toward a more localized configuration with enhanced out-of-plane Ru-$d{z^2}$ character, increased O-$p_{\pi}$ density, and sizable Ru moments.

This density-based picture provides a microscopic complement to the variational trend in Fig.~\ref{fig:E_vs_U}. 
In the same parameter window where the AFM trial state develops a sizable Ru moment, we observe a clear reorganization of Ru-O hybridization rather than a weak spin polarization of an otherwise unchanged itinerant state. 
This also helps explain why compressive strain shifts the variational optimum toward larger $\omega$: once the lattice is distorted toward a more magnetic regime, the lower-energy trial states acquire more of the localized Ru-O character visualized in Fig.~\ref{fig:charge_diff}(b). 
At the same time, Fig.~\ref{fig:E_vs_U} shows that such large-moment AFM trial states are variationally penalized in the unstrained system, indicating that the preference for the NM state is determined by energetics rather than by the absence of an allowed altermagnetic texture. 

These differences in charge-density persist even at higher values of $\omega$ where the system is driven into an insulating state. This suggests that RuO$_2$ is also close to a metal-insulator phase transition, and similar to other bad-metals such as VO$_2$ or ABO$_3$-perovskite metals, as charge increasingly accumulate in the O-$p$ orbitals due to chemical doping it can be driven to a charge- and bond-disproportionated insulating phase~\cite{ABO3_MIT,VO2_bad_metal}. Given the very small number of known insulating altermagnets, our study paves the way for rationally designing more such altermagnetic insulators. 

\textit{Conclusion.—}
The Fixed-node DMC establishes the stoichiometric rutile RuO$_2$ as nonmagnetic in the pristine bulk limit, resolving the functional-dependent mean-field controversy.
Within the PBE0 trial family, the variational optimum occurs at $\omega=0.10$, where the nonmagnetic state lies $23(9)\ \mathrm{meV/f.u.}$ below the lowest AFM state considered.
This result reconciles recent $\mu$SR evidence for the absence of a static bulk moment~\cite{hiraishi2024nonmagnetic,kessler2024absence} with earlier theoretical reports of AFM order, and places RuO$_2$ not in an intrinsically ordered bulk phase, but on the nonmagnetic side of a nearby magnetic instability.

At the same time, we find that epitaxial-like uniaxial strain along $z$ strongly reshapes this delicate balance by shifting the optimal $\omega$ and stabilizing magnetism under compression.
In particular, $3\%$ compression drives RuO$_2$ into an AFM state with an ordered moment of $0.97(2)$ $\mu_{B}$ in DMC.
This pronounced strain tunability provides a natural framework for understanding earlier altermagnetic-like signatures~\cite{feng2022anomalous,bose2022tilted,fedchenko2024observation,zhu2019anomalous,berlijn2017itinerant} and ongoing reports of altermagnetic responses in epitaxial RuO$_2$ films~\cite{he2025evidence,jeong2025anisotropic,jeong2025metallicity,forte2025strain,brahimi2025confinement,lytvynenko2026magnetic,akashdeep2026surface} where anisotropic strains can persist, without invoking an intrinsically magnetic stoichiometric bulk ground state. Further, a rational strategy for making an insulating altermagnetic phase emerges from this study by systematically increasing charge- and bond-disproportionation in the RuO$_2$ structure.

More broadly, our results show that in itinerant materials near magnetic instabilities, where density-functional predictions can vary qualitatively with functional choice and tuning parameters, fixed-node DMC can provide a decisive many-body benchmark for resolving disputed magnetic energetics.

\textit{Acknowledgements.—}
This work was primarily supported by the U.S. Department of Energy, Office of Science, Basic Energy Sciences, Materials Sciences and Engineering Division, as part of the Computational Materials Sciences Program and Center for Predictive Simulation of Functional Materials. 
J. Ahn (initial calculations and analysis), P. Ganesh (mentorship, analysis, writing) and J. T. Krogel (mentorship, analysis, writing) were supported by the U.S. Department of Energy, Office of Science, Basic Energy Sciences, Materials Sciences and Engineering Division, as part of the Computational Materials Sciences Program and Center for Predictive Simulation of Functional Materials.
J.Ahn also acknowledges support from U.S. Department of Energy, Office of Science, Office of Basic Energy Sciences, Computational Materials Sciences Award No. DE-SC0020177 for final calculations, analysis and writing of the paper. 
S.-H.Kang (concept, analysis, and writing) was supported by Korea Institute of Science and Technology Information (KISTI) (K26L4M2C2-01).
An award of computer time was provided by the Innovative and Novel Computational Impact on Theory and Experiment (INCITE) program. This research used resources of the Argonne Leadership Computing Facility, which is a DOE Office of Science User Facility supported under contract DE-AC02-06CH11357. This research also used resources of the Oak Ridge Leadership Computing Facility, which is a DOE Office of Science User Facility supported under Contract DE-AC05-00OR22725.

This manuscript has been authored by UT-Battelle, LLC under Contract No. DE-AC05-00OR22725 with the U.S. Department of Energy. The United States Government retains and the publisher, by accepting the article for publication, acknowledges that the United States Government retains a non-exclusive, paid-up, irrevocable, worldwide license to publish or reproduce the published form of this manuscript, or allow others to do so, for United States Government purposes. The Department of Energy will provide public access to these results of federally sponsored research in accordance with the DOE Public Access Plan (http://energy.gov/downloads/doe-public-access-plan).

\bibliographystyle{apsrev4-2}
\bibliography{references}

\begin{thebibliography}{44}
\bibitem [{not()}]{note}%
  \BibitemOpen
  \bibinfo {note} {$\omega$ denotes the exact-exchange mixing fraction (often denoted $\alpha$), not a range-separation parameter.}\BibitemShut {Stop}%
\end{thebibliography}

\end{document}